# THE PHYSICS OF MICROWAVE BACKGROUND ANISOTROPIES [†]

Wayne Hu,[1] Naoshi Sugiyama[1,2], & Joseph Silk[1]

[1]*Department of Astronomy and Physics and Center for Particle Astrophysics*

[2]*Department of Physics, Faculty of Science The University of Tokyo, Tokyo 113, Japan*

Cosmic microwave background anisotropies provide a vast amount of information on both structure formation in the universe and the background dynamics and geometry. The full physical content and detailed structure of anisotropies can be understood in a simple and intuitive fashion through a systematic investigation of the individual mechanisms for anisotropy formation, based on elementary gravitational and fluid dynamics.

hu@pac2.berkeley.edu

> *Words are for catching ideas; once you've caught the idea, you can forget about the words. Where can I find a man who knows how to forget about words so that I might have a few words with him?*
>
> *–Chuang-tzu*

With the COBE DMR measurement of anisotropies at large angles[1] and the rapidly growing number of detections at degree scales,[2] the enormous wealth of information stored in the cosmic microwave background (CMB) is becoming available. Anisotropies bear the imprint, filtered through the dynamics and geometry of the expanding universe, of the fluctuations which eventually form structure in the universe. In general then, they encode information on both the model for structure formation in the universe and the background cosmology. It is important to study the process of anisotropy formation in order to separate and understand these factors.

Many physical effects together form rather complicated structures in the anisotropy spectrum.[3,4,5,6] Though complete, solutions from numerical integration of radiation transport equations leave their physical content opaque to all but the specialist.[7,8,9] It is the goal of this article to separate anisotropy formation into its component mechanisms. We concentrate here on extracting the essential properties in a physically intuitive manner and refer the interested reader to our more technical treatments for details.[10,11] Features in the anisotropy power spectrum, along with their dependence on the background cosmology,[10,11,12,13,14] can be directly traced to individual physical processes. Their robustness to model changes provides realistic hope that fundamental cosmological parameters will be measured from the CMB.

## Anisotropy Formation

Fluctuations in the total matter density, which includes decoupled species such as the neutrinos and possibly collisionless dark matter, interact with the photons through the gravitational potentials they create. These same fluctuations grow by gravitational attraction, *i.e.* infall into their own potential wells, to eventually form large scale structure in the universe. Their presence in the early universe is also responsible for anisotropy formation.

Before redshift $z_* \simeq 1000$, the CMB was hot enough to ionize hydrogen. Compton scattering off electrons, which are in turn linked to the protons through Coulomb interactions, strongly couples the photons to the baryons and establishes a photon-baryon fluid. Photon pressure resists compression of the fluid by gravitational infall and sets up acoustic oscillations. At $z_*$, recombination forms neutral hydrogen and the photons last scatter. Regions of compression and rarefaction at this epoch represent hot and cold spots respectively. Photons also suffer gravitational redshifts from climbing out of the potentials on the last scattering surface. The resultant fluctuations appear to the observer today as anisotropies on the sky. By developing this simple picture in greater detail, we show how realistic anisotropies such as those depicted in Fig. 1 are formed.

The equations of motion for temperature fluctuations take on a simple form when decomposed into normal modes. These are plane waves for a flat geometry, referred to here as such even when considering their open geometry generalization.[15,16] We represent temperature fluctuations in Newtonian form, which simplifies concepts such as infall and redshift, by defining them on the spatial hypersurfaces of the conformal Newtonian gauge.[17,18]

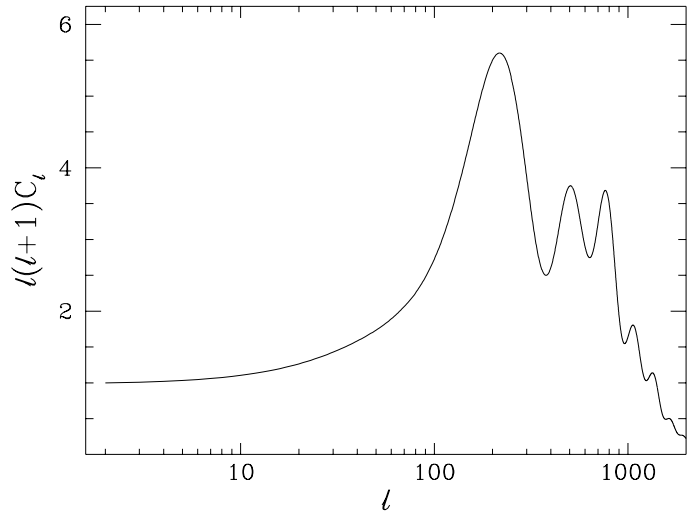

**Figure 1.** The anisotropy power spectrum $C_\ell$ as a function of multipole number $\ell$ in the standard cold dark matter model with $\Omega_0 = 1$, $h = 0.5$, $\Omega_B = 0.05$, scale invariant scalar initial fluctuations, and arbitrary normalization. The corresponding angle on the sky is approximately $100/\ell$ degrees.

Under the gravitational force $F$, a temperature perturbation $\Theta_0 = \Delta T/T$ of comoving wavenumber $k$ evolves almost as a simple harmonic oscillator before recombination,[10] $\ddot{\Theta}_0 + k^2 c_s^2 \Theta_0 \simeq F$. The overdots represent derivatives with respect to conformal time $\eta = \int (1+z) dt$ and the sound speed $c_s$, which measures the resistance of the fluid to compression, is related to the speed of light as $c_s \equiv \dot{p}/\dot{\rho} = c/\sqrt{3(1+R)}$, where the baryon-photon density ratio is proportional to $R = 3\rho_b/4\rho_\gamma = 3.0 \times 10^4 (1+z)^{-1} \Omega_B h^2$. Here the Hubble constant is given by $H_0 = 100 h \,\mathrm{km\,s^{-1}\,Mpc^{-1}}$ and $\Omega_B$ is the fraction of the critical density in baryons.

Gravity drives the oscillator with a force $F = -k^2 c^2 \Psi/3 - \ddot{\Phi}$, where $\Psi$ is the Newtonian gravitational potential, obtained from density fluctuations via the generalized Poisson equation, and $\Phi \simeq -\Psi$ is the perturbation to the space curvature. They also represent plane wave fluctuations in the time-time and space-space metric components respectively. The sign convention reflects the fact that overdensities create positive space curvature and negative potentials, *i.e.* potential wells. In real space though, a single plane wave represents both overdense *and* underdense regions. We use the former to guide intuition since the distinction is only in sign.

**Acoustic Oscillations: Basics**

Let us first consider temperature fluctuations before recombination in the case of a *static* potential.[19,20,21] Although only appropriate for a universe which has *always* been matter dominated, it illustrates the general nature of the acoustic oscillations. In this case, $F = -k^2 c^2 \Psi/3$ and represents the usual driving force of gravity that leads to infall into potential wells. Since big bang nucleosynthesis implies that the baryon density is low, $\Omega_B h^2 \simeq 10^{-2}$, as a first approximation, assume that the photons completely dominate the fluid $c_s \simeq c/\sqrt{3}$.

Gravitational infall compresses the fluid until resistance from photon pressure reverses the motion. Since the gravitational force is constant in this case, it merely shifts the zero point of the oscillation to $\Theta_0 = -\Psi$. We are however free to choose the initial conditions, assumed here to be adiabatic:[3,10] $\Theta_0(0) = -\frac{2}{3}\Psi$ and $\dot{\Theta}_0(0) = 0$ (see Fig. 2a). In this case, the photons follow the total matter, making the temperature higher inside a potential well. The effective initial displacement of $\Theta_0(0) + \Psi = \frac{1}{3}\Psi$ then evolves as $\Theta_0(\eta) = \frac{1}{3}\Psi \cos(kc_s\eta) - \Psi$. At last scattering $\eta_*$, the photons decouple from the baryons and stream out of potential wells suffering gravitational redshifts equal to $\Psi$. We thus call $\Theta_0 + \Psi$ the *effective* temperature fluctuation. Here the redshift exactly cancels the zero point displacement since gravitational infall and redshift are one and the same for a photon-dominated system.

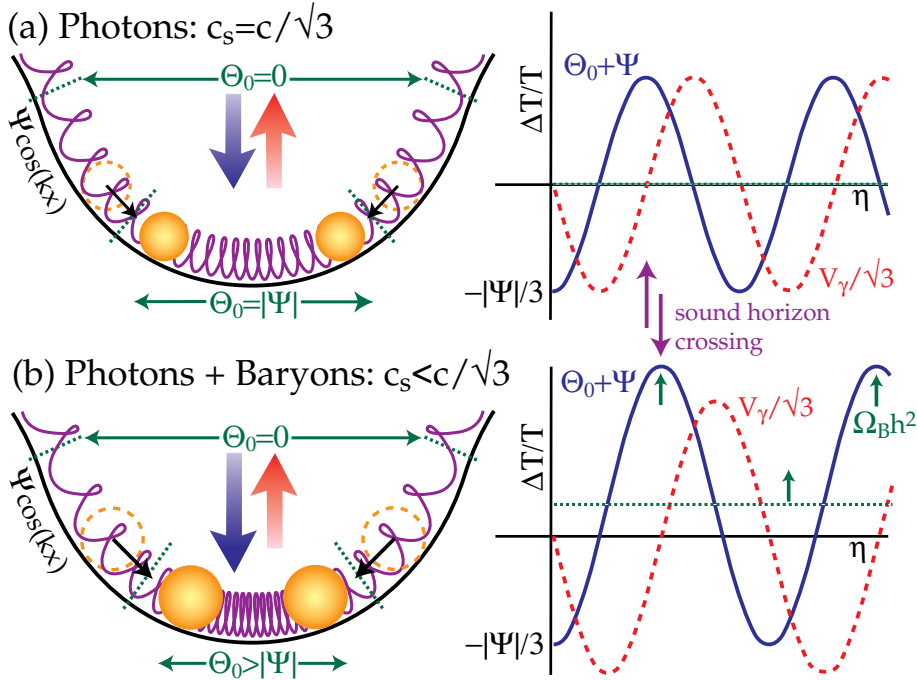

**Figure 2.** Acoustic Oscillations. (a) Photon-dominated system. Fluid compression through gravitational infall is resisted by photon pressure setting up acoustic oscillations. Displayed here is a potential well in real space $-\pi/2 \lesssim kx \lesssim \pi/2$. Gravity displaces the zero point so that at the bottom of the well, the temperature is $\Theta_0 = |\Psi| = -\Psi$ at equilibrium with $\Psi/3$ excursions. This displacement is exactly cancelled by the redshift $\Psi$ a photon experiences climbing out from the bottom of the potential well. Velocity oscillations lead to a Doppler effect 90 degrees phase shifted from the temperature perturbation. (b) Photon-baryon system. Baryons increase the gravitating mass, causing more infall and a net zero point displacement, even after redshift. Temperature crests (compression) are enhanced over troughs (rarefaction) and velocity contributions.

The phase of the oscillation at last scattering determines the effective fluctuation. Since the oscillation frequency $\omega = kc_s$, the critical wavenumber $k = \pi/c_s\eta_*$ is essentially at the scale of the *sound horizon* $c_s\eta_*$ (see Fig 2). Larger wavelengths will not have evolved from the initial conditions and possess $\frac{1}{3}\Psi$ fluctuations after gravitational redshift. This combination of the intrinsic temperature fluctuation and the gravitational redshift is the well known Sachs-Wolfe effect.[3] Shorter wavelength fluctuations can be frozen at different phases of the oscillation. Since fluctuations as a function of $k$ go as $\cos(kc_s\eta_*)$ at last scattering, there will be a harmonic series of temperature *fluctuation* peaks with $k_m = m\pi/c_s\eta_*$ for the $m$th peak. Odd peaks thus represent the compression phase (temperature crests), whereas even peaks represent the rarefaction phase (temperature troughs), inside the potential wells.

**Displaced Oscillations: Baryon Content**

Though effectively pressureless, the baryons still contribute to the inertial and gravitational mass of the fluid. This decreases the sound speed and changes the balance of pressure and gravity. Gravitational infall now leads to greater compression of the fluid in a potential well, *i.e.* a further displacement of the oscillation zero point (see Fig. 2b). Since the redshift is not affected by the baryon content, this relative shift remains after last scattering to enhance all peaks from compression over those from rarefaction. If the baryon photon ratio $R$ were constant, $\Theta(\eta) + \Psi = \frac{1}{3}\Psi(1+3R)\cos(kc_s\eta) - R\Psi$, with compressional peaks a factor of $(1+6R)$ over the $R=0$ case. In reality, the effect is reduced since $R \to 0$ at early times.

Finally the *evolution* of the sound speed has a minor effect on its own. In classical mechanics, the ratio of energy to frequency of an oscillator $\omega$ is an adiabatic invariant. Thus for the slow changes in $\omega \propto c_s$, the amplitude of the oscillation varies as $\Theta_0 \propto c_s^{1/2} \propto (1+R)^{-1/4}$. Since $R(\eta_*) = 30\Omega_B h^2 \lesssim 1$ at recombination, this is ordinarily not a strong effect.

**Velocity Oscillations: Doppler Effect**

Since the turning points are at the extrema, the fluid velocity oscillates 90 degrees out of phase with the density (see Fig. 2b). Its motion relative to the observer causes a Doppler shift. Whereas the observer velocity creates a pure dipole anisotropy on the sky, the fluid velocity causes a spatial temperature variation $V_\gamma/\sqrt{3}$ on the last scattering surface from its line of sight component. For a photon-dominated $c_s \simeq c/\sqrt{3}$ fluid, the velocity contribution is equal in amplitude to the density effect.[19,21] This photon-intrinsic Doppler shift should be distinguished from the scattering-induced Doppler shift of reionized scenarios.[6]

The addition of baryons significantly changes the relative velocity contribution. As the effective mass increases, conservation of energy requires that the velocity decreases for the same initial temperature displacement. Thus the *relative* amplitude of the velocity scales as $c_s$. In the toy model of a constant baryon-photon density ratio $R$, the oscillation becomes $V_\gamma/\sqrt{3} = \frac{1}{3}\Psi(1+3R)(1+R)^{-1/2}\sin(kc_s\eta)$. Notice that velocity oscillations are symmetric around zero leading to an even more prominent compressional peaks (see Fig. 2b). Even in a universe with $\Omega_B h^2$ given by nucleosynthesis, $R$ is sufficiently large to make velocity contributions subdominant.

**Driven Oscillations: Radiation Content**

All realistic models involve potentials which are time-dependent, leading to a non-trivial gravitational driving force that can greatly enhance the prominence of the acoustic peaks.[10,11] We have hitherto assumed that matter dominates the energy density. In reality, radiation dominates above the redshift of equality $z_{eq} = 2.4 \times 10^4 \Omega_0 h^2$, assuming the usual three flavors of massless neutrinos. The feedback from radiation perturbations into the gravitational potential makes the CMB sensitive to the matter-radiation ratio in the background *and* the fluctuations.

Consider first adiabatic initial conditions as before. Inside the sound horizon, radiation pressure prevents gravitational infall during radiation domination. Energy density fluctuations consequently can no longer maintain a constant gravitational potential. Counterintuitively, this decaying potential can actually enhance temperature fluctuations through its near resonant driving force. Since the potential decays after sound horizon crossing, it mimics $\cos(kc_s\eta)$ for $kc_s\eta \lesssim \pi$. Consequently, it drives the first compression without a counterbalancing effect on the subsequent rarefaction or gravitational redshift.

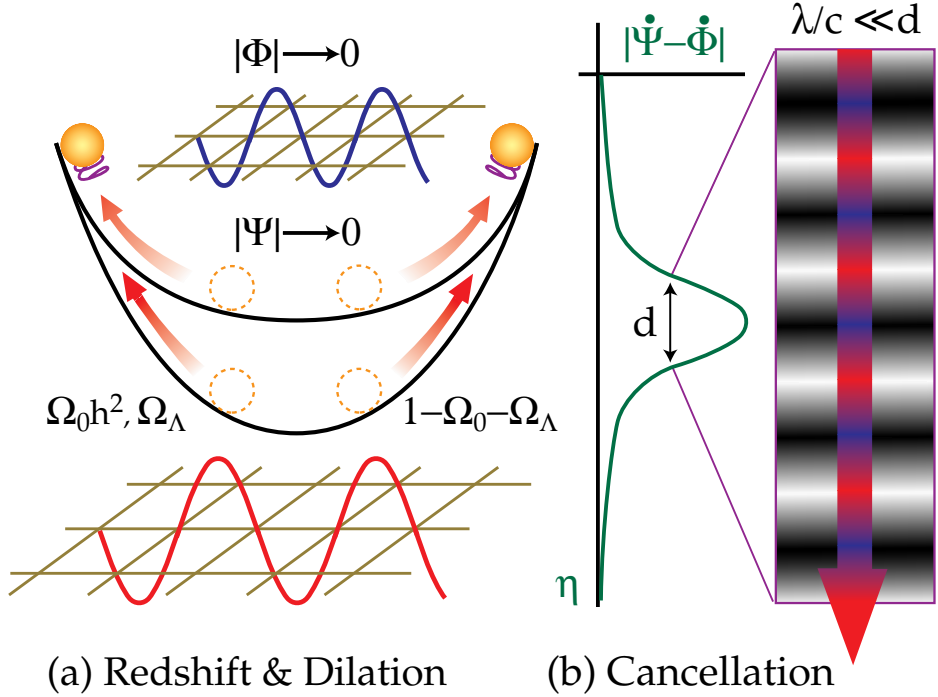

(a) Redshift & Dilation     (b) Cancellation

**Figure 3.** Gravitational redshift and dilation effects in a time dependent potential. Time variability occurs whenever the matter is not the sole dynamical factor and thus probes $\Omega_0 h^2$, $\Omega_\Lambda$, $1 - \Omega_0 - \Omega_\Lambda$ and any isocurvature perturbations. (a) Decay of the potential $|\Psi|$ decreases the gravitational redshift leading to an effective blueshift in the well. The implied curvature perturbation $|\Phi|$ decay represents a "contraction of space" which blueshifts photons through time dilation, nearly doubling the $\Psi$ effect. (b) In the free streaming limit after last scattering, these two mechanisms combine to form the ISW effect. Redshift-blueshift cancellation cuts off contributions at small scales where the photon traverses many wavelengths during the decay.

Moreover, there is another effect. Recall that the space curvature perturbation follows the potential as $\Phi \simeq -\Psi$. Since the forcing function $F = -\ddot{\Phi} - k^2 c^2 \Psi/3$, a changing $\Phi$ also drives oscillations. As $\Phi$ is a perturbation to the spatial metric, its change induces a time-dilation effect which is wholly analogous to the cosmological redshift due to the expansion. Heuristically, the overdensities which establish the potential well "stretch" the space-time fabric (see Fig. 3a). As the potential well decays, it re-contracts. Photons which are caught in this contraction find their wavelength similarly contracted, i.e. blueshifted. Thus a differential change in $\Phi$ leads to a dilation effect, $\dot{\Theta}_0 = -\dot{\Phi}$, and consequently a forcing effect on $\ddot{\Theta}_0$ of $-\ddot{\Phi}$ as required.

If $\Psi$ were exactly $\cos(k c_s \eta)$, then $\ddot{\Phi}$ would double the driving force. In reality, the oscillation amplitude is boosted to $\simeq \frac{3}{2}\Psi(0)$.[11] Only short wavelengths, which cross the sound horizon during the radiation dominated epoch, experience this enhancement. For the popular $\Omega_0 h^2 \simeq 0.25$ models, the sound horizon at equality is several times smaller than that at last scattering. Hence delaying equality, by lowering $\Omega_0 h^2$ or increasing the number of relativistic species, boosts the amplitude of oscillations for the first few peaks. Finally, the decay of the potential $\Psi$ also removes the zero point shift and thus lifts the pattern of alternating heights for the peaks.

As a second example of forced oscillations, consider isocurvature perturbations. In this case, the matter alone carries the initial fluctuations, i.e. $\Theta_0(0) = 0$ and since the radiation dominates the energy density, $\Phi(0) = 0 = \Psi(0)$ as well. However $\dot{\Theta}(0) \neq 0$ and is set to counteract the gravitational attraction of the matter. Consequently, the potential grows to be significant only near sound horizon crossing and subsequently decreases if the universe is radiation dominated. The forcing function resembles $\sin(kc_s\eta)$ and thus drives the sine harmonic of oscillations. Furthermore, since fluctuations are initially established to counter gravity, infall enhances *even* rather than odd peaks. Outside the sound horizon, dilation implies that $\Theta_0(\eta_*) = -\Phi(\eta_*)$, creating a Sachs-Wolfe effect of $[\Theta_0 + \Psi](\eta_*) \simeq 2\Psi(\eta_*)$.

**Damped Oscillations: Photon Diffusion**

In reality, the photons and baryons are not perfectly coupled since the photons possess a mean free path in the baryons $\lambda_C$ due to Compton scattering. As the photons random walk through the baryons, hot spots and cold spots are mixed. Fluctuations thereafter remain only in the unscattered fraction causing a near exponential decrease in amplitude as the diffusion length $\lambda_D \sim \sqrt{N}\lambda_C = \sqrt{c\eta\lambda_C}$ overtakes the wavelength.[4]

At last scattering, the ionization fraction $x_e$ decreases due to recombination, thus increasing the mean free path of the photons $\lambda_C \propto (x_e n_b)^{-1}$. The effective diffusion scale is therefore extremely sensitive to the ionization history in addition to the baryon number density $n_b$. Subtle effects during and even before last scattering can have a measurable effect on the damping.[22,23] Moreover, if last scattering is delayed, *e.g.* by early reionization, diffusion continues and can destroy all the acoustic peaks. Assuming a standard recombination history however, the approximate scaling can be obtained from the Saha equation for the ionization at fixed redshift or temperature, $x_e \propto (\Omega_B h^2)^{-1/2}$. The final damping length therefore approximately scales as $\lambda_D(\eta_*) \propto \eta_*^{1/2}(\Omega_B h^2)^{-1/4}$.

**Free Streaming: ISW Effect**

After last scattering, the photons free stream toward the observer. Only gravitational effects can further alter the temperature. The differential redshift from $\dot{\Psi}$ and dilation from $\dot{\Phi}$ discussed above must be integrated along the trajectory of the photons. We thus call the combination the *integrated* Sachs-Wolfe (ISW) effect.[3] For adiabatic models, it can contribute via the potential decay for modes that cross the sound horizon between last scattering and full matter domination. In isocurvature models, potential *growth* outside the sound horizon makes the ISW effect dominate over the Sachs-Wolfe effect for all wavelengths larger than the sound horizon at $\eta_*$. Because these effects are sensitive to the radiation content and occur primarily at early times, we call them *early* ISW effects.

One additional subtlety is introduced in ISW effects. Since the photons free stream, they can travel across many wavelengths of the perturbation during the decay of the potential. If the potential decays while the photon is in an underdense region, it will suffer an effective redshift rather than a blueshift. Contributions from overdense and underdense regions will cancel and damp the ISW effect if the decay time is much greater than the light travel time across a wavelength (see Fig. 3b). The damping does not occur for the *early* ISW effect. Since it arises when the perturbations are outside or just crossing the horizon, the time scale for the decay is always less than, or comparable to, the light travel time across a wavelength.

In an open or $\Lambda$ model, the universe enters a rapid expansion phase once matter no longer dominates the expansion. We call the effect of the resultant potential decay the *late* ISW effect. Decay takes on the order of an expansion time at curvature or $\Lambda$ domination independent of the wavelength. Thus, cancellation leads to a gradual damping in $k$ of contributions as the wavelength becomes smaller than the horizon at the decay epoch. For a fixed $\Omega_0$, the decay epoch occurs much later in flat $\Omega_\Lambda + \Omega_0 = 1$ models than open ones where $\Omega_\Lambda = \Lambda c^2/3H_0^2$. Thus $\Lambda$ models will suffer cancellation of late ISW contributions at a much larger scale than open models.[24] In summary, the epoch that the universe exits the radiation ($\Omega_0 h^2$) and matter dominated phase ($\Omega_\Lambda, 1 - \Omega_0 - \Omega_\Lambda$) is imprinted on the CMB by the early and late ISW effects respectively.

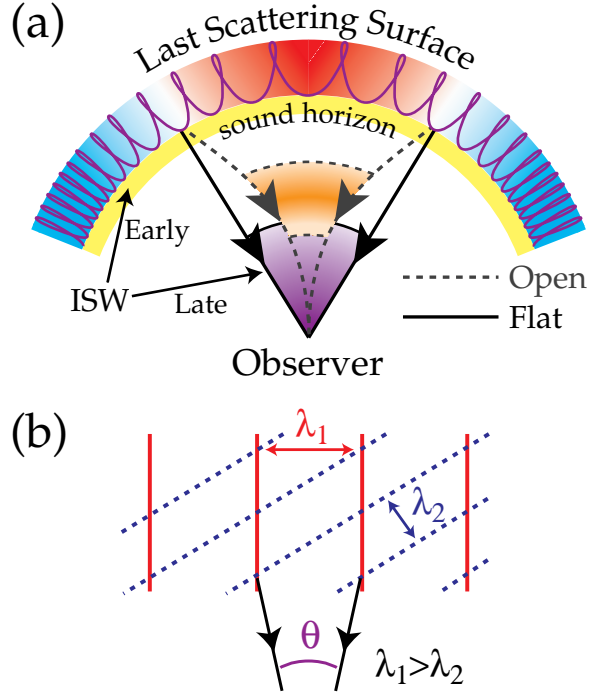

**Figure 4.** Projection effects. (a) Acoustic contributions exhibit a series of peaks with decreasing angle beginning at the angular scale the sound horizon subtends at last scattering. This scale decreases significantly as the curvature increases due to geodesic deviation. Contributions after last scattering, come from a smaller physical scale for the same angular scale, which pushes the late ISW effect of flat $\Lambda$ and open models to larger angles. (b) The orientation of the plane wave projected on the surface of last scattering leads to aliasing of power from shorter wavelengths onto larger angles. This smooths out sharp features and prevents a steeply rising (blue) anisotropy spectrum.

**Angle-Distance Conversion: Projection Effects**

We have been considering the generation of temperature fluctuations in space. However, what one actually observes are temperature anisotropies on the sky. The connection between the two is that a spatial fluctuation on a distant surface, say at last scattering for the acoustic effects, appears as an anisotropy on the sky. Three quantities go into this conversion: the spectrum of spatial fluctuations, the distance to the surface of their generation, and curvature or lensing in light propagation to the observer (see Fig. 4a).

For the acoustic contributions, the $k$ modes that reach extrema in their oscillation at last scattering form a harmonic series of peaks related to the sound horizon. This in turn is approximately $c\eta_*/\{1 + C[1 + R(\eta_*)]^{1/2}\}$, where $R(\eta_*) = 30\Omega_B h^2$ and $C \simeq \sqrt{3} - 1$. Since $\Omega_B h^2$ must be low to satisfy nucleosynthesis constraints, the sound horizon will scale roughly as the particle horizon $c\eta_*$. The particle horizon at last scattering itself scales as $c\eta_* \propto (\Omega_0 h^2)^{-1/2} f_R$. Here $f_R = [1 + (24\Omega_0 h^2)^{-1}]^{1/2} - (24\Omega_0 h^2)^{-1/2}$ and is near unity if the universe is matter-dominated at $\eta_*$. In a flat $\Lambda$ universe, the distance to the last scattering surface scales approximately as $c\eta_0 \propto (\Omega_0 h^2)^{-1/2} f_\Lambda$ with $f_\Lambda = 1 + 0.085 \ln \Omega_0$. Notice that the two behave similarly at high $\Omega_0 h^2$. Since the acoustic angle $\theta_A \propto \eta_*/\eta_0$, the leading term $(\Omega_0 h^2)^{-1/2}$ has no effect. Slowly varying corrections from $f_R/f_\Lambda$ decreases the angular scale somewhat as $\Omega_0 h^2$ is lowered. On the other hand, the damping scale subtends an angle $\theta_D \simeq \lambda_D/\eta_0 \propto (\Omega_0 h^2)^{1/4}(\Omega_B h^2)^{-1/4} f_R^{1/2}/f_\Lambda$. Even in a low $\Omega_0 h^2$ universe $\theta_D$ is only weakly dependent on $h$ unlike $\theta_A$ the acoustic scale.

By far the most dramatic effect is due to *background* curvature in the universe.[25] If the universe is open, photons curve on their geodesics such that a given scale subtends a much smaller angle in the sky than in a flat universe. In a $\Lambda = 0$ universe, the angle-distance relation yields $\theta_A \propto \eta_*\Omega_0 h$, *i.e.* $\propto \Omega_0^{1/2} f_R$. Likewise, the damping scale subtends an angle $\theta_D \propto \lambda_D \Omega_0 h$, *i.e.* $\propto \Omega_0^{3/4}\Omega_B^{-1/4} f_R^{1/2}$. At *asymptotically* high and low $\Omega_0 h^2$, $f_R \simeq 1$ and $f_R \propto (\Omega_0 h^2)^{1/2}$ respectively, so that there is a weak but different scaling with $h$ and strong but similar scaling with $\Omega_0$ for the two angles. The latter should be an easily measureable effect.[26]

Contributions from after last scattering, such as the ISW effects, arise from a distance closer to us. A given scale thus subtends a larger angle on the sky (see Fig. 4). Their later formation also implies that the radiation correction factor $f_R$ will be smaller. For example, the angle subtended by the adiabatic early ISW effect scales nearly as $\Omega_0^{1/2}$ in a $\Lambda = 0$ universe even at low $\Omega_0 h^2$.

The above discussion implicitly assumes an one-to-one correspondence of linear scale onto angle that is strictly only true if the wavevector is perpendicular to the line of sight. In reality, the orientation of the wavevector leads to aliasing of different, in fact larger, angles for a given wavelength (see Fig. 4b). This is particularly important for Doppler contributions which vanish for the perpendicular mode.[10] Moreover if there is a lack of long wavelength power, *e.g.* in typical baryon isocurvature models, large angle anisotropies are dominated by aliasing of power from short wavelengths. Consequently, the angular power spectrum may be less blue than the spatial power spectrum.[11] On the other hand, for so called "scale invariant" or equal weighting of $k$ modes, aliasing tends to smear out sharp features but does not change the general structure of the real to angular space mapping. It is evident that gravitational lensing from the curvature *fluctuations* of overdense and underdense regions has a similar but usually smaller effect.[27]

## Anisotropy Spectrum

We have shown that anisotropy formation is a simple process that is governed by gravitational effects on the photon-baryon fluid and the photons alone before and after last scattering respectively. The component contributions contain detailed information on classical cosmological parameters. Let us now put them together to form the total anisotropy spectrum.

### An Adiabatic Example

The popular scale invariant adiabatic models provide a useful example of how cosmological information is encoded into the anisotropy spectrum. Specifically by scale invariant, we mean that the logarithmic contribution to the gravitational potential is initially constant in $k$. For open universes, this is only one of several reasonable choices near the curvature scale.[28,29,30,31] In Fig. 5, we display a schematic representation of the anisotropy spectrum which separates the various effects discussed above and identifies their dependence on the background cosmology.

Changing the overall dynamics from $\Omega_0 = 1$ through flat $\Lambda$ models to open models is similar to shifting the spectrum in angular space toward smaller angles. Beginning at the largest angles, the ISW effect from late potential decay dominates in $\Omega_0 \ll 1$ models. Cancellation suppresses contributions for wavelengths smaller than the particle horizon at the exit from matter domination. This damping extends to larger angles in $\Lambda$ than in open models affecting even the quadrupole. At scales much larger than the sound horizon at $\eta_*$ and particle horizon at equality, the effective temperature, or Sachs-Wolfe effect, is $[\Theta + \Psi](\eta_*) \simeq \frac{1}{3}\Psi(\eta_*)$. Shifting equality through $\Omega_0 h^2$ changes the redshift contribution $\Psi(\eta_*)$. For scales just above the sound horizon, the early ISW effect boosts fluctuations as the radiation content is increased by lowering $\Omega_0 h^2$. In sufficiently low $\Omega_0$ open models, the late and early ISW effects merge and entirely dominate over the last scattering surface effects at large angles.

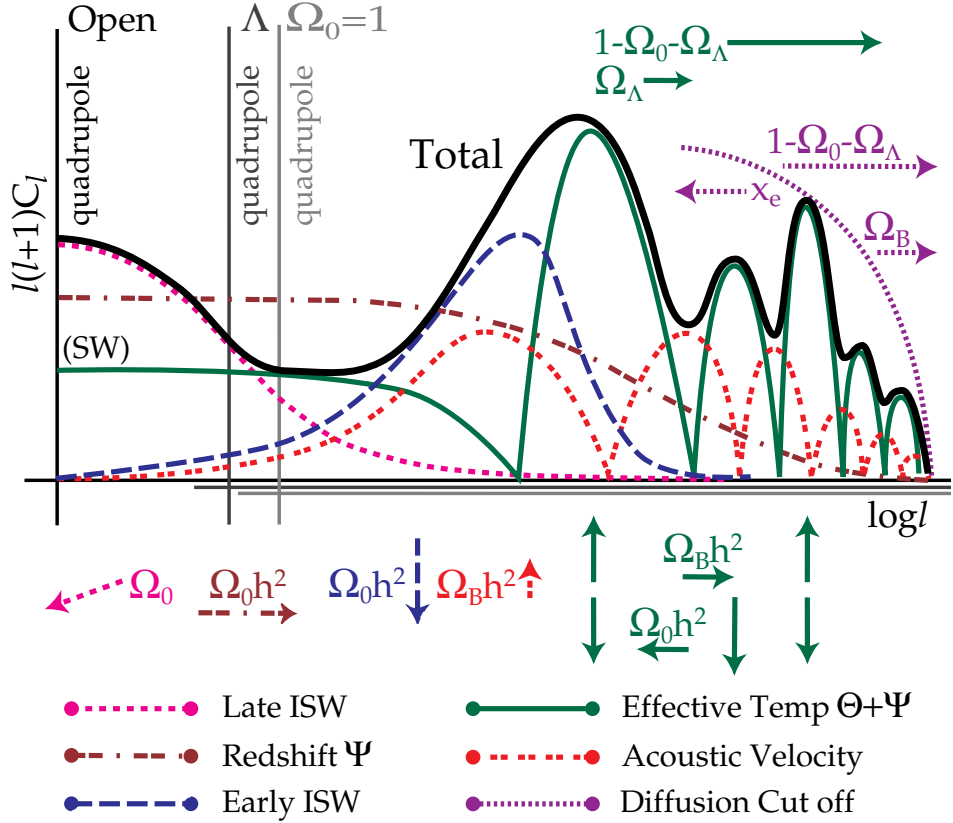

**Figure 5.** Total anisotropy power spectrum in a schematic representation for scale invariant adiabatic scalar models. Features in open models are shifted to significantly smaller angles compared with $\Lambda$ and $\Omega_0 = 1$ models, represented here as a shift in the $\ell$ axis beginning at the quadrupole $\ell = 2$. The monopole and dipole fluctuations are unobservable due to the mean temperature and peculiar velocity at the point of observation. The effective temperature at last scattering $[\Theta + \Psi](\eta_*)$ includes the gravitational redshift effect $\Psi(\eta_*)$. At large scales, the effective temperature goes to $\Psi(\eta_*)/3$ and is called the Sachs-Wolfe (SW) contribution. In reality, small scale acoustic contributions from the effective temperature and velocity are smoothed out somewhat in $\ell$ due to projection effects[10] (Fig. 4b).

The first of a series of peaks from the acoustic oscillations appear on the sound horizon at $\eta_*$. In the total spectrum, the first acoustic peak merges with the early ISW effect. A lower $\Omega_0 h^2$ thus serves to broaden out and change the angular scaling of this combined feature. The acoustic peak heights also depend strongly on $\Omega_0 h^2$ for the first few peaks due to the driving effects of infall and dilation. Furthermore, greater infall due to the baryons allows more gravitational zero point shifting if $\Omega_0 h^2$ is sufficiently high to maintain the potentials. Odd peaks will thus be enhanced over the even, as well as velocity contributions, with increasing $\Omega_B h^2$. The location of the peaks is dependent on the sound horizon, distance to last scattering, and the curvature. In a low $\Omega_B h^2$, high $\Omega_0 h^2$ universe, it is sensitive only to the curvature $1 - \Omega_0 - \Omega_\Lambda$. Finally, the physics of recombination sets the diffusion damping scale which cuts off the series of acoustic peaks.

**Robustness**

How robust are anisotropies to model changes? Obviously, changing the initial spectrum will significantly modify the spectrum. For example, isocurvature conditions and tilt can alter the relative contributions of the various effects. The lack of super-curvature modes in open inflationary models can also suppress the low order multipoles.[32] On the other hand, they may be boosted by gravitational wave ISW contributions.[33,34]

Acoustic oscillations however are unavoidable, if there are potential perturbations before last scattering. Even exotic models such as defect-induced fluctuations should give rise to oscillations. Since adiabatic and isocurvature conditions drive two different harmonics, they can be distinguished by the relation between the peaks and the sound horizon at last scattering.[11] The *locations* of the peaks are then dependent only on the background cosmology, *i.e.* mainly on the curvature but also on a combination of $\Omega_B h^2$, $\Omega_\Lambda$ and $\Omega_0 h^2$. On the other hand, the difference in heights between odd and even peaks is a reasonably robust probe of the baryon-photon ratio, *i.e.* $\Omega_B h^2$, relative to the matter-radiation ratio at last scattering, *i.e.* $\Omega_0 h^2$ and possibly even the number of massless neutrinos. Finally, the damping scale probes the baryon content and the detailed physics of recombination. If acoustic oscillations are detected in the anisotropy data, clearly we will be able to measure many parameters of classical cosmology.

The one caveat to these considerations is that reionization can completely erase the acoustic oscillations. In a model with sufficiently early reionization, *i.e.* $z_i \gg 10$, the photon diffusion length grows to be the horizon scale at the new last scattering surface and consequently damps all the peaks. Even in this pessimistic case, the CMB will yield important information on the first generation of structure in the universe through the location and detailed behavior of the damping regime. With the high precision measurements expected from the next generation of anisotropy experiments, an abundance of cosmological information will undoubtably be available to us.

**Acknowledgments**

We would like to thank E. Bunn, D. Scott, and M. White for useful discussions. This work was supported by the NSF and NASA.